\documentclass{article}
\usepackage[utf8]{inputenc} 
\usepackage[T1]{fontenc}    
\usepackage{hyperref}       
\usepackage{url}            
\usepackage{booktabs}       
\usepackage{amsfonts}       
\usepackage{nicefrac}       
\usepackage{microtype}      
\usepackage{color}
\usepackage{times}
\usepackage{epsfig}
\usepackage{graphicx}
\usepackage{amsmath}
\usepackage{amssymb}
\usepackage{algorithm}
\usepackage{multirow}
\usepackage{makecell}
\usepackage{colortbl}
\usepackage{tabularx}
\usepackage{bbm}
\usepackage{algorithmic}
\usepackage{enumitem}
\usepackage{pifont}         
\usepackage{wrapfig}
\usepackage{subcaption}
\newcolumntype{L}[1]{>{\raggedright\arraybackslash}p{#1}}
\newcolumntype{C}[1]{>{\centering\arraybackslash}p{#1}}
\newcolumntype{R}[1]{>{\raggedleft\arraybackslash}p{#1}}
\setlist[itemize]{leftmargin=4.mm}

\usepackage[preprint]{corl_2022} 

\title{Forecasting Cryptocurrency Staking Rewards}

\author{
  Sauren Gupta\thanks{The contributions of Sauren Gupta to this work were made during his tenure as an intern at Coinbase.}, 
  Apoorva Hathi Katharaki\thanks{Corresponding author. Email: \texttt{apoorva.hathi@coinbase.com}}, 
  Yifan Xu, 
  Bhaskar Krishnamachari\thanks{Dr. Bhaskar Krishnamachari is a paid consultant for Coinbase and has contributed to this paper in this capacity.}, 
  Rajarshi Gupta \\
  Machine Learning Team, Coinbase \\
}

\begin{document}
\maketitle


\begin{abstract}
This research explores a relatively unexplored area of predicting cryptocurrency staking rewards, offering potential insights to researchers and investors. We investigate two predictive methodologies: a) a straightforward sliding-window average, and b) linear regression models predicated on historical data. The findings reveal that ETH staking rewards can be forecasted with an RMSE within 0.7\% and 1.1\% of the mean value for 1-day and 7-day look-aheads respectively, using a 7-day sliding-window average approach. Additionally, we discern diverse prediction accuracies across various cryptocurrencies, including SOL, XTZ, ATOM, and MATIC. Linear regression is identified as superior to the moving-window average for perdicting in the short term for XTZ and ATOM. The results underscore the generally stable and predictable nature of staking rewards for most assets, with MATIC presenting a noteworthy exception.
\end{abstract}


\section{Introduction}
	
Staking, within the context of cryptocurrency, is a mechanism wherein holders of specific digital currencies, such as Ethereum or others utilizing a proof-of-stake (PoS) model, earn rewards by actively engaging in transaction validation on a blockchain while they continue to hold on to their assets \cite{saleh2021blockchain, buterin2013proof}. In essence, staking involves dedicating a specified quantity of cryptocurrency assets to fortify a blockchain network, subsequently facilitating the verification of transactions. Contrasting with proof-of-work (PoW) systems, where computational power is crucial for transaction verification and block generation, a PoS model entrusts validation to stakers \cite{nair2021evaluation, buterin2013proof}. These participants willingly lock a predetermined quantity of the network’s currency as collateral. Stakers, in turn, are selected—often proportionally to their staked amount—to validate transactions and introduce new blocks to the blockchain \cite{nguyen2019proof}. Engaging in staking not only allows individuals to earn rewards for strengthening the network but also augments their cryptocurrency holdings, thus offering an enticing incentive to procure and retain the cryptocurrency, thereby enhancing its overall security and stability \cite{nguyen2019proof}.

Many digital currency networks, like ETH, Cosmos, Tezos, Algorand, Cardano, and Solana, give rewards to users who agree to lock up ("stake") their digital tokens for a set time \cite{buterin2013ethereum, kwon2019cosmos}. This is an incentive for users and it also makes the network more secure. For platforms that manage digital currencies and blockchain networks, staking can also make customers more loyal because it creates a kind of attachment or "stickiness."

The rewards allocated for staking are subject to change, contingent upon various factors such as the number of participants staking, the maturity of the project, prevailing market conditions, among others \cite{john2021equilibrium}. This inherent variability and unpredictability in the rewards offered for staking cryptocurrencies pose a multifaceted challenge. Various entities that invest, encompassing individual investors, institutional investors, and platforms managing digital assets, may find predictive information regarding future staking rewards beneficial. This could assist them in determining optimal times to stake or un-stake their assets within a specific network. 

The solution we propose in the paper involves the application of predictive models to forecast the staking rewards of a variety of cryptocurrencies. Our approach is predicated on the assumption that historical trends and patterns in staking rewards can provide valuable insights into future performance. Two predictive approaches have been considered in this context, sliding-window average and linear regression. Sliding-window average essentially involves computing the mean rewards over a specified time window, which is a straightforward and intuitive method that can provide a quick snapshot of the trend in staking rewards. However, it is important to note that this method assumes that the future will mirror the past, which may not always be the case, particularly in the volatile and unpredictable world of cryptocurrencies. Linear regression is an effective predictive model that can capture more complex patterns in the data. It assumes a linear relationship between the independent and dependent variables, which may not always hold true. However, with careful feature selection and model tuning, linear regression can provide robust and reliable predictions. While more sophisticated models such as LSTM \citep{hochreiter1997long} could be used for time-series prediction, we have limited ourselves to these simpler models because they already show good performance, as the results show. Despite extensive research in the field of cryptocurrency, there is a noticeable gap in the literature regarding the prediction of staking rewards. This study aims to fill this gap by exploring this under-researched area. By employing straightforward but efficient predictive models, this research seeks to provide insights into the predictability of staking rewards across various cryptocurrencies. The findings of this study could potentially provide valuable data for stakeholders and investors, aiding them in making informed staking decisions.

Our study shows that the sliding-window average method is effective in predicting Ethereum (ETH) staking rewards, with a Root Mean Square Error (RMSE) of 0.7\% for a 1-day forecast and less than 1.1\% for a 7-day forecast.This method surprisingly surpasses linear regression models in predicting staking rewards for this specific asset. However, for cryptocurrencies such as Tezos (XTZ) and Cosmos (ATOM), linear regression models proved to be more effective than the sliding-window average method in predicting staking rewards. The effectiveness of the sliding-window average and linear regression models was assessed by calculating the RMSE between the actual and predicted staking rewards over different forecast periods (1-day and 7-days). Additionally, the performance of these models was evaluated across various cryptocurrencies (ETH, SOL, XTZ, ATOM, and MATIC), offering a comprehensive view of their applicability and efficiency. The findings of this study reveal a diverse range of predictive accuracy across various cryptocurrencies. In particular, the sliding-window average model exhibits a strong predictive capacity for Ethereum (ETH). Conversely, Tezos (XTZ) and Cosmos (ATOM) demonstrate a higher forecasting proficiency when utilizing linear regression models for short term. A common trend across the majority of assets is the formation of slow-varying and moderately predictable time series in staking rewards. However, Polygon (MATIC) deviates from this pattern, displaying notably erratic rewards patterns.

In summary, our study consists of the following highlights:
\begin{itemize}
    \item Application of ML techniques for accurate staking rewards predictions across multiple assets, with remarkable precision for short-term and long-term forecasts.
    \item Demonstrated good performance of simple models like averaging historical rewards and linear regression for cryptocurrency staking rewards forecasting.
\end{itemize}

\section{Related Works}
There has been prior research work exploring various aspects of staking of cryptocurrencies. For example, \citet{choi2023optimal} formulate an optimal staking problem in which they explore tradeoffs between rewards and illiquidity. \citet{john2021equilibrium} analyze equilibrium staking levels and show that staking levels may not always be increasing in block rewards. \citet{xu2022reap} provide a general survey of rewards farming in DeFi including considerations of staking as one of the components of rewards farming. Some of this research has focused specifically on staking pools and concerns about centralization.  For instance, \citet{he2020staking} explore the question of staking centralization and conditions under which a stable equilibrium exists for staking pools; and \citet{gersbach2022staking} model the formation of staking pools in the presence of malicious adversaries.  To our knowledge the research literature on staking has not previously explored the problem of forecasting staking rewards, the focus of this paper. 

While there has been decades of research on forecasting financial time series \cite{tang2022survey}, in particular stock prices \citep{soni2022machine}, over the past decade there has also been a lot of prior work on modeling and forecasting cryptocurrency prices and the volatility of their prices. \citet{kyriazis2020systematic} present a survey of modeling the “bubble” dynamics of cryptocurrency prices. \citet{khedr2021cryptocurrency} and \citet{amirzadeh2022applying} present a survey of applications of artificial intelligence and machine learning to cryptocurrency price and volatility prediction. This literature has explored a number of different machine learning models for predictions. For example, \citet{mudassir2020time} explore ML-based time-series forecasting of Bitcoin prices using high-dimensional features. \citet{hamayel2021novel} consider the use of LSTM models for cryptocurrency price prediction. \citet{derbentsev2021comparative} present a comparative study of different ML prediction algorithms. \citet{d2022deep} present application of deep learning to predict the volatility of cryptocurrencies.

Our work adds a new dimension to the above literature on application of machine learning methods to predictions in the context of cryptocurrencies by exploring staking rewards as the focus of the predictions. We find that in this work that in contrast to the high volatility of cryptocurrency prices which has been the primary focus of prior work, staking rewards are relatively slower-moving and simpler time series, and consequently, relatively simpler ML models show good performance. 

\section{Background: Stacking and rewards}
Staking is a fundamental concept in the realm of cryptocurrencies, allowing users to participate in a network’s operations while receiving rewards. It's essential to understand that different platforms have varied approaches and rules governing their staking processes, particularly in how the staking rewards are determined. Here's a general breakdown:

\subsection{How Staking Works}
Staking involves users, or "stakers," holding and locking up a cryptocurrency in a wallet to participate in the network’s consensus mechanism, typically Proof of Stake (PoS) or its variants. By doing this, stakers help secure the network, validate transactions, and maintain the blockchain’s overall integrity. In return, they receive staking rewards, usually in the form of cryptocurrency.

\subsection{Determining rewards}
Rewards from staking can be influenced by various factors, depending on the specific network’s protocol. Generally, the more tokens held and the longer they are staked, the higher the potential return. Factors influencing rewards may include the total amount of cryptocurrency staked in the network, staking duration, token inflation rate, and network transaction fees.

\subsection{Staking Duration}
Different platforms might have varied rules regarding the length of time the cryptocurrencies need to be held for staking ("staking duration") and the time it takes to withdraw them ("un-staking time"). Some networks may require a minimum staking duration to qualify for rewards, and there might be a waiting period (cooling-off period) for accessing or "unstaking" the staked tokens. This approach encourages network stability by discouraging rapid, speculative movements of funds.

\section{Feature Selection}

In our study, we tried to understand the details of staking rewards, focusing on the following cryptocurrencies: ETH, SOL, XTZ, MATIC and ATOM. We carefully chose and reviewed various datasets and features. It was crucial to pick data that was helpful and of good quality, considering that there is limited data available for some specific assets. This section explains the datasets we used, the features we looked at, and the thoughtful methods behind leaving out certain features.

\subsection{Considered Features}

Table \ref{tab:feature_descriptions} below delineates the selected features and their corresponding data sources that were utilized in this study. These features were extracted from a variety of sources, including Coinbase data storage, Google Trends and YFinance. The raw data obtained from these sources underwent significant structuring and refinement to tailor it for analysis.

\begin{table}[h]
\centering
\caption{Overview of Extracted Features and Their Data Sources}
\label{tab:feature_descriptions}
\begin{tabular}{|l|l|}
\hline
\textbf{Feature}          & \textbf{Description}                                          \\ \hline
Time series of rewards    & Historical data representing rewards over a period of time    \\ \hline
Price feed                & Real-time price information of the asset                      \\ \hline
Trends         & The frequency of searches for the cryptocurrency asset        \\ \hline
\end{tabular}
\end{table}

\begin{figure}[h]
\centering

\begin{subfigure}[b]{0.3\textwidth}
  \centering
  \includegraphics[width=\linewidth]{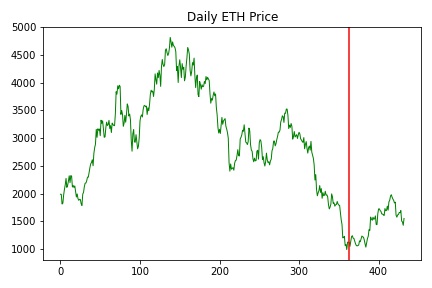}
  \caption{Daily ETH price}
  \label{fig:sub1_1}
\end{subfigure}%
\hfill
\begin{subfigure}[b]{0.3\textwidth}
  \centering
  \includegraphics[width=\linewidth]{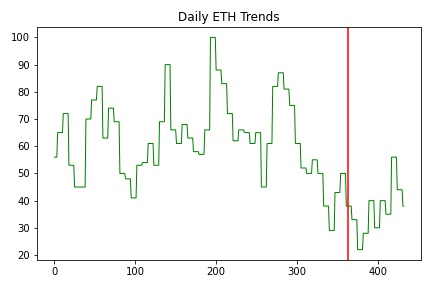}
  \caption{ETH trends}
  \label{fig:sub1_2}
\end{subfigure}%
\hfill
\begin{subfigure}[b]{0.3\textwidth}
  \centering
  \includegraphics[width=\linewidth]{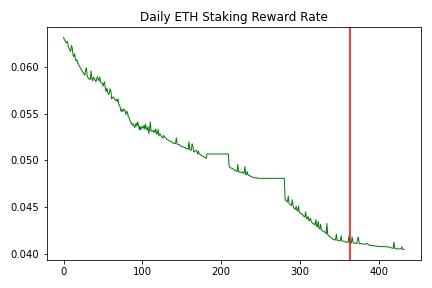}
  \caption{ETH reward rate}
  \label{fig:sub1_3}
\end{subfigure}

\caption{Ethereum (ETH) Data Visualization (2021-06-23 to 2022-08-06) - Showcasing Daily Price, Search Trends, and Reward Rate.}
\label{fig:feature_plot}
\end{figure}

Taking the example of ETH, a collection of time-series data for about one year from 2021-06-23 to 2022-08-06 was diligently gathered in Figure \ref{fig:feature_plot}. This data, related to the features mentioned in Table \ref{tab:feature_descriptions} , was closely examined to understand its relationship and influence on staking rewards.

\subsection{Exclusion and Unexplored Features}

In our study, we also looked at some additional features to understand their importance and effect, which are mentioned in Table \ref{table:exclusion_features}. After careful review, some features were left out after thoughtful consideration and testing. Specific features such as ‘staking volume’ and ‘percentage of staked volume’ were excluded post-analysis, as they were found to closely determine the staking rewards for ETH. The similarity and near redundancy of these features, upon normalization, further justified their omission. Moreover, post-merge values for ETH were deliberately left out due to observed variations in trends and a lack of substantial data to ensure a fortified model training and testing process.

Certain features, despite being considered, were not incorporated into the final model. These unexplored features include: User count, Distribution of staked tokens, Transaction data on the asset. The decision to not utilize these features was a calculated one, ensuring that only the most significant and influential features were utilized, optimizing the study’s objective of analyzing staking rewards.

\begin{table}[h]
\centering
\caption{Excluded and unexplored features}
\label{table:exclusion_features}
\begin{tabular}{|l|p{9cm}|}
\hline
\textbf{Feature}                    & \textbf{Description}                                  \\ \hline
Staking volume                      & Total amount staked on the asset                      \\ \hline
Percentage of staked volume         & Percentage of total assets that are staked            \\ \hline
Distribution of staked tokens       & Granularity based on addresses/wallets                \\ \hline
Transaction data on asset           & Detailed transaction data associated with the asset   \\ \hline
User count                          & Number of users engaged in staking activities         \\ \hline
\end{tabular}
\end{table}

\section{Experimental Setup}

\subsection{Forecasting Algorithms}
For forecasting the staking rewards, two algorithms were employed: the moving-window average and linear regression models. Here is a brief overview of the methods used:

\begin{itemize}
    \item \textbf{Moving-Window Average} (MWA): The algorithm involves using a set number of past data points to calculate an average that predicts future values. Due to its simplicity and absence of an explicit training/learning process, it serves as a fundamental test, assuming that the rewards changes relatively slowly.
    \item \textbf{Linear Regression}: In this approach, simple linear regression was utilized to enhance the model’s predictive accuracy. Simple linear regression involves creating a linear relationship between a single explanatory variable and the dependent variable. This method is used to forecast future values based on past trends, and it's a common practice to improve the performance of forecasting models.
    Two approaches of linear regression were employed 
    \begin{itemize} 
    	\item \textbf{Single Linear Regression} (SLR): We use a single feature -  as input to the linear regression model.
    	\item \textbf{Multiple Linear Regression} (MLR): We use multiple features as input to the multiple regression model.
    \end{itemize}
\end{itemize}
\subsection{Objective of the Model}
Our objective was to develop a machine learning model adept at forecasting the staking rewards of ETH and other stakeable assets for a specified number of upcoming days (n). The experimentation started with forecasts for the next day (n=1), gradually extending to a week ahead (n=7), to assess and compare the performance and reliability of the results obtained.
\subsection{Data Collection and Utilization}
Data for 2021-06-23 to 2022-08-06 was assembled, focusing on features pivotal to staking rewards. A comparison was conducted between linear regression models and moving-average model, aiming to gauge the efficacy of the staking rewards prediction models for n-day ahead forecasts.
\subsection{Training and Test Data Splitting Strategy}
An approach was devised for splitting the data into training and test sets. Various strategies were explored to ascertain the most effective method of partitioning the data to optimize the model’s performance in forecasting. We found that training for three previous months and using it to test for the next month gives good performance. This informed approach allowed for the refinement of the model to achieve a more insightful and dependable forecasting setup.
\subsection{Metrics}
The metrics considered  is root mean square error (RMSE), which measures the average difference between a statistical model's predicted values and the actual values.

\section{Experimental Results}

\subsection{Staking Rewards Next Day Prediction}
\begin{table}[h]
\centering
\caption{Staking Rewards Next Day Prediction Performance (RMSE/Mean)}
\label{table:regression_models}
\begin{tabular}{|l|p{2.7cm}|p{3.4cm}|p{4cm}|}
\hline
\textbf{Asset} & Moving-Window Average & Single Linear Regression (rewards only) & Multiple Linear Regression (rewards, price, trends) \\ \hline
ETH (pre-merge) & 0.007 & 0.007 & 0.009 \\ \hline
SOL             & 0.029 & 0.031 & 0.041 \\ \hline
MATIC           & 0.610 & 0.673 & 0.807 \\ \hline
ATOM            & 0.017 & 0.011 & 0.018 \\ \hline
XTZ             & 0.054 & 0.046 & 0.101 \\ \hline
\end{tabular}
\end{table}

In Table \ref{table:regression_models}, we present staking rewards prediction for 5 assets with Metrics: RMSE divided by the mean
\begin{itemize}
    \item Moving-Window Average: we predict the next day’s rewards as the average of the past 7 days.
    \item Linear Regression with single feature (rewards only) and multiple features (rewards, price, trends) Here we consider 7 days data to predict the next day, ie, we feed in 7 days of data as input and predict the next day’s value. And Train-Test Split: 90 days for training followed by 30 days for test. 
\end{itemize}
For ETH, moving window average and single feature linear regression perform comparably. For ATOM and XTZ, single feature linear regression outperforms MWA and multiple linear regression. For SOL and MATIC, MWA performs the best among all methods. MATIC performs poorly compared to other tokens with all the methods, which seems to be a consequence of the highly volatile rewards rate for the token.

\begin{figure}[h]
\centering

\begin{subfigure}[b]{0.3\textwidth}
  \centering
  \includegraphics[width=\linewidth]{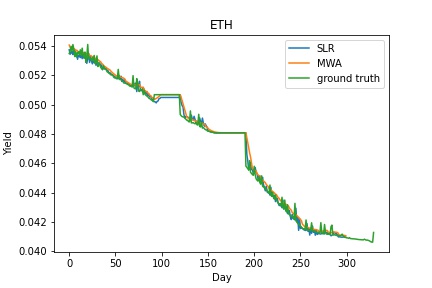}
  \caption{ETH}
  \label{fig:sub2_1}
\end{subfigure}%
\hfill
\begin{subfigure}[b]{0.3\textwidth}
  \centering
  \includegraphics[width=\linewidth]{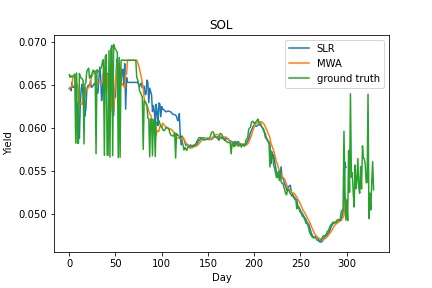}
  \caption{SOL}
  \label{fig:sub2_2}
\end{subfigure}
\hfill
\begin{subfigure}[b]{0.3\textwidth}
  \centering
  \includegraphics[width=\linewidth]{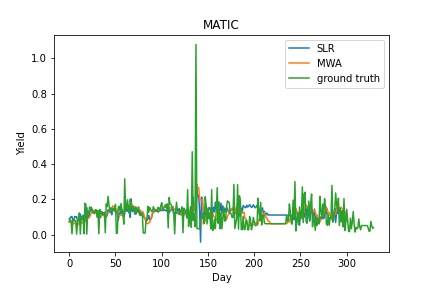}
  \caption{MATIC}
  \label{fig:sub2_3}
\end{subfigure}%

\begin{subfigure}[b]{0.3\textwidth}
  \centering
  \includegraphics[width=\linewidth]{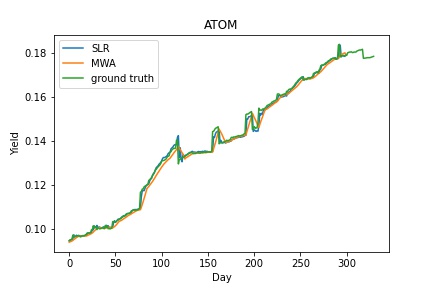}
  \caption{ATOM}
  \label{fig:sub2_4}
\end{subfigure}%
\hfill
\begin{subfigure}[b]{0.3\textwidth}
  \centering
  \includegraphics[width=\linewidth]{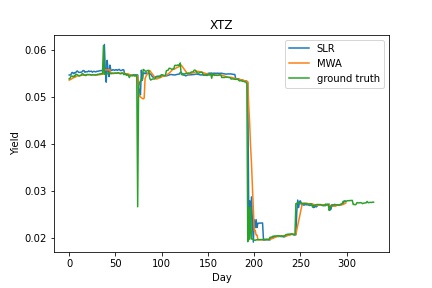}
  \caption{XTZ}
  \label{fig:sub2_5}
\end{subfigure}

\caption{Staking Rewards Next Day Predictions of rewards for 5 cryptocurrencies ETH, SOL, MATIC, ATOM and XTZ, using MWA and SLR}
\label{fig:predictions_1day}
\end{figure}
In Figure \ref{fig:predictions_1day}, we show the plotted predictions of all the tokens with the chosen parameters from Table \ref{table:regression_models}, for 316 days from 2021-09-14 to 2022-07-27. The plots show the predictions of different methods - moving window average, single linear regression and multiple linear regression. SOL and MATIC show high volatility in actual rewards during certain times.

\subsection{Staking Rewards Next N-Days Prediction}
We further explore the prediction of rewards when the n-days are predicted instead of just 1 day ahead prediction, as shown in Table \ref{table:errors_ndays}, for 316 days from 2021-09-14 to 2022-07-27. Metrics are RMSE divided by the mean. The moving window average performs better overall for all of the tokens except MATIC. SLR performs better than MWA for shorter term prediction like 1-day or 2-day ahead.

\begin{figure}[h]
	\centering
	
	\begin{subfigure}[b]{0.3\textwidth}
		\centering
		\includegraphics[width=\linewidth]{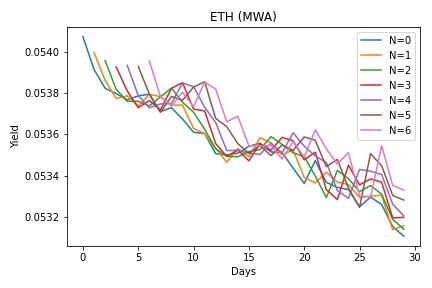}
		\caption{ETH}
		\label{fig:sub_mwa_eth}
	\end{subfigure}%
	\hfill
	\begin{subfigure}[b]{0.3\textwidth}
		\centering
		\includegraphics[width=\linewidth]{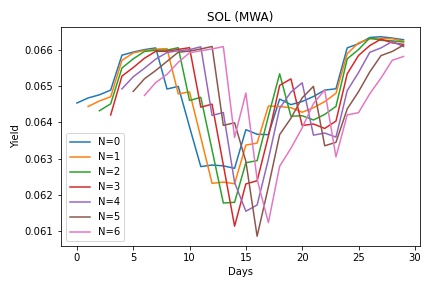}
		\caption{SOL}
		\label{fig:sub_mwa_sol}
	\end{subfigure}
	\hfill
	\begin{subfigure}[b]{0.3\textwidth}
		\centering
		\includegraphics[width=\linewidth]{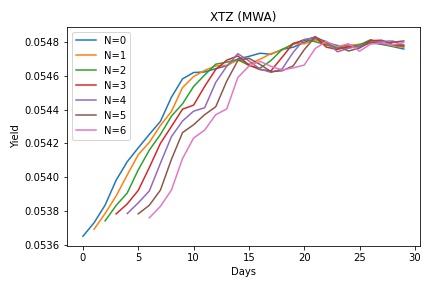}
		\caption{XTZ}
		\label{fig:sub_mwa_xtz}
	\end{subfigure}%
 
	\begin{subfigure}[b]{0.3\textwidth}
		\centering
		\includegraphics[width=\linewidth]{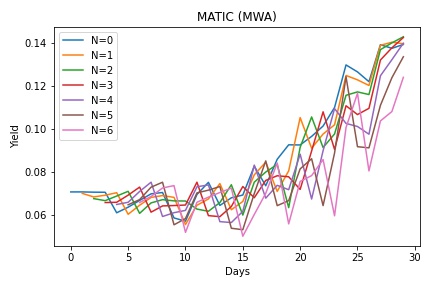}
		\caption{MATIC}
		\label{fig:sub_mwa_matic}
	\end{subfigure}%
	\hfill
	\begin{subfigure}[b]{0.3\textwidth}
		\centering
		\includegraphics[width=\linewidth]{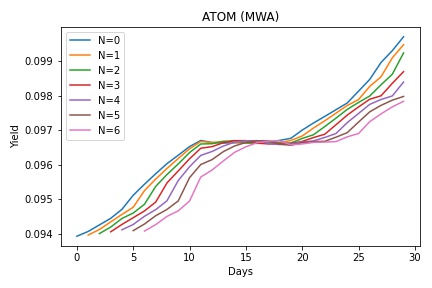}
		\caption{ATOM}
		\label{fig:sub_mwa_atom}
	\end{subfigure}
	
\caption{Staking Rewards Next N-day Predictions of 5 cryptocurrencies ETH, SOL, MATIC, ATOM and XTZ using MWA}
\label{fig:predictions_mwa_ndays}
\end{figure}

\begin{figure}[h]
	\centering
	
	\begin{subfigure}[b]{0.3\textwidth}
		\centering
		\includegraphics[width=\linewidth]{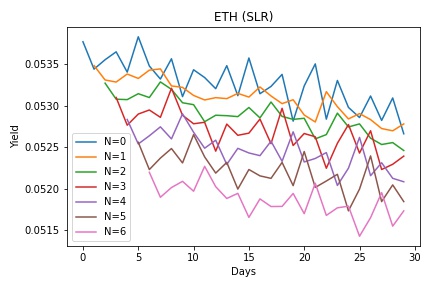}
		\caption{ETH}
		\label{fig:sub_ln_eth}
	\end{subfigure}%
	\hfill
	\begin{subfigure}[b]{0.3\textwidth}
		\centering
		\includegraphics[width=\linewidth]{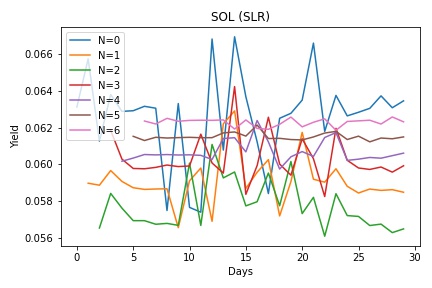}
		\caption{SOL}
		\label{fig:sub_ln_sol}
	\end{subfigure}
	\hfill
	\begin{subfigure}[b]{0.3\textwidth}
		\centering
		\includegraphics[width=\linewidth]{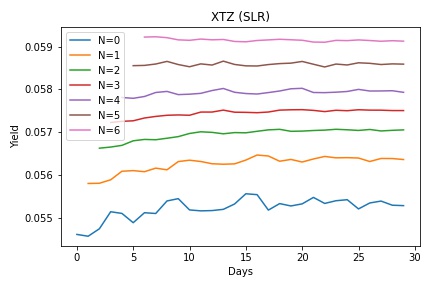}
		\caption{XTZ}
		\label{fig:sub_ln_xtz}
	\end{subfigure}%
 
	\begin{subfigure}[b]{0.3\textwidth}
		\centering
		\includegraphics[width=\linewidth]{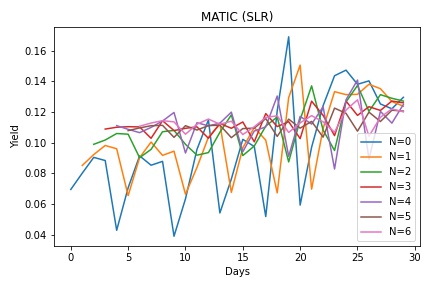}
		\caption{MATIC}
		\label{fig:sub_ln_matic}
	\end{subfigure}%
	\hfill
	\begin{subfigure}[b]{0.3\textwidth}
		\centering
		\includegraphics[width=\linewidth]{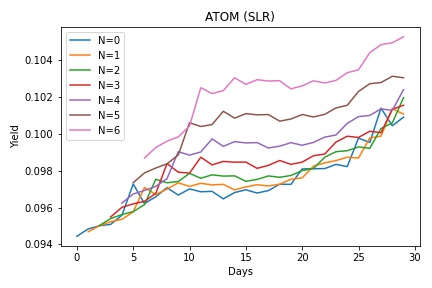}
		\caption{ATOM}
		\label{fig:sub_ln_atom}
	\end{subfigure}
	
	\caption{Staking Rewards Next N-day Predictions of 5 cryptocurrencies ETH, SOL, MATIC, ATOM and XTZ using SLR}
	\label{fig:predictions_linear_ndays}
\end{figure}

In Figure \ref{fig:predictions_mwa_ndays} and Figure \ref{fig:predictions_linear_ndays}, we show the plotted predictions of rewards all tokens for 1 to 7 days ahead, with the chosen parameters from Table \ref{table:errors_ndays} for MWA and SLR approach. The predictions plotted are for the first 30 days of our datatset. The plots show that the predictions are more precise for MWA, although the fluctuations are better captured by the linear regression model.
  
\begin{table}[h!]
\centering
\caption{Staking Rewards Next N-Day Prediction Performance Over Days (RMSE/Mean)}
\label{table:errors_ndays}
\begin{tabular}{c|cc|cc|cc|cc|cc}
\toprule
 & \multicolumn{2}{c|}{ETH (pre-merge)} & \multicolumn{2}{c|}{SOL} & \multicolumn{2}{c|}{XTZ} & \multicolumn{2}{c|}{MATIC} & \multicolumn{2}{c}{ATOM} \\
N & MWA & SLR & MWA & SLR & MWA & SLR & MWA & SLR & MWA & SLR \\
\midrule
1 & \textbf{0.007} & \textbf{0.007} & \textbf{0.029} & 0.032 & 0.051 & \textbf{0.048} & \textbf{0.612} & 0.628 & 0.017 & \textbf{0.011} \\
2 & \textbf{0.008} & \textbf{0.008} & \textbf{0.023} & 0.041 & \textbf{0.056} & 0.061 & \textbf{0.613} & 0.623 & 0.019 & \textbf{0.017} \\
3 & \textbf{0.008} & 0.009 & \textbf{0.031} & 0.048 & \textbf{0.061} & 0.079 & \textbf{0.619} & 0.639 & \textbf{0.020} & 0.023 \\
4 & \textbf{0.009} & 0.011 & \textbf{0.034} & 0.051 & \textbf{0.065} & 0.097 & 0.634 & \textbf{0.630} & \textbf{0.022} & 0.030 \\
5 & \textbf{0.010} & 0.014 & \textbf{0.034} & 0.054 & \textbf{0.070} & 0.114 & 0.655 & \textbf{0.639} & \textbf{0.024} & 0.036 \\
6 & \textbf{0.010} & 0.018 & \textbf{0.037} & 0.061 & \textbf{0.075} & 0.130 & 0.663 & \textbf{0.635} & \textbf{0.026} & 0.042 \\
7 & \textbf{0.011} & 0.022 & \textbf{0.040} & 0.071 & \textbf{0.083} & 0.145 & 0.671 & \textbf{0.634} & \textbf{0.028} & 0.048 \\
\bottomrule
\end{tabular}
\end{table}

\section{Discussion and Conclusion}
\vspace{-2mm}
We find that among all our strategies, 7-day Moving Window Average for one-day predictions gives between 0.7-5\% error (RMSE/MEAN) and performs well for all tokens, except MATIC. Prediction for MATIC performs poorly compared to other tokens with all the methods, which seems to be a consequence of the highly volatile rewards rate for the token. 

When looking at short term prediction, i.e., n<3, for ETH, MWA and single feature linear regression perform comparably. For ATOM and XTZ, single feature linear regression outperforms moving window average and multiple linear regression. For SOL and MATIC, moving window average performs the best among all methods.

For longer days ahead prediction, i.e., n>=3, MWA outperforms other methods for all tokens except MATIC. This shows that a simple approach such as moving window average is sufficient and the effectiveness of complicated ML algorithms reduces as we start predicting further days ahead into the future.

For all assets except MATIC, a 7-days ahead prediction using MWA has an increased error of no more than 3.2\% compared to a 1-day ahead prediction.

Simple linear regression based on the staking rewards time series alone was found to outperform multiple linear regression using other features. This shows the effectiveness of simpler strategies when it comes to prediction of staking rewards.

Some questions that could be explored in future work include handling of non-stationary token data such as pre and post-merge ETH and when to retrain them.

\textbf{Acknowledgement}
Sauren Gupta’s involvement in this paper reflects work that he did while he was employed as an intern at Coinbase in 2022. Dr. Krishnamachari is a paid consultant for Coinbase and has assisted on this paper in this capacity.

\bibliography{stacking}  

\begin{thebibliography}{21}
\providecommand{\natexlab}[1]{#1}
\providecommand{\url}[1]{\texttt{#1}}
\expandafter\ifx\csname urlstyle\endcsname\relax
  \providecommand{\doi}[1]{doi: #1}\else
  \providecommand{\doi}{doi: \begingroup \urlstyle{rm}\Url}\fi

\bibitem[Saleh(2021)]{saleh2021blockchain}
F.~Saleh.
\newblock Blockchain without waste: Proof-of-stake.
\newblock \emph{The Review of financial studies}, 34\penalty0 (3):\penalty0
  1156--1190, 2021.

\bibitem[Buterin(2013)]{buterin2013proof}
V.~Buterin.
\newblock What proof of stake is and why it matters.
\newblock \emph{Bitcoin Magazine}, 26, 2013.

\bibitem[Nair and Dorai(2021)]{nair2021evaluation}
P.~R. Nair and D.~R. Dorai.
\newblock Evaluation of performance and security of proof of work and proof of
  stake using blockchain.
\newblock In \emph{2021 Third International Conference on Intelligent
  Communication Technologies and Virtual Mobile Networks (ICICV)}, pages
  279--283. IEEE, 2021.

\bibitem[Nguyen et~al.(2019)Nguyen, Hoang, Nguyen, Niyato, Nguyen, and
  Dutkiewicz]{nguyen2019proof}
C.~T. Nguyen, D.~T. Hoang, D.~N. Nguyen, D.~Niyato, H.~T. Nguyen, and
  E.~Dutkiewicz.
\newblock Proof-of-stake consensus mechanisms for future blockchain networks:
  fundamentals, applications and opportunities.
\newblock \emph{IEEE access}, 7:\penalty0 85727--85745, 2019.

\bibitem[Buterin et~al.(2013)]{buterin2013ethereum}
V.~Buterin et~al.
\newblock Ethereum white paper.
\newblock \emph{GitHub repository}, 1:\penalty0 22--23, 2013.

\bibitem[Kwon and Buchman(2019)]{kwon2019cosmos}
J.~Kwon and E.~Buchman.
\newblock Cosmos whitepaper.
\newblock \emph{A Netw. Distrib. Ledgers}, 27, 2019.

\bibitem[John et~al.(2021)John, Rivera, and Saleh]{john2021equilibrium}
K.~John, T.~J. Rivera, and F.~Saleh.
\newblock Equilibrium staking levels in a proof-of-stake blockchain.
\newblock \emph{Available at SSRN 3965599}, 2021.

\bibitem[Hochreiter and Schmidhuber(1997)]{hochreiter1997long}
S.~Hochreiter and J.~Schmidhuber.
\newblock Long short-term memory.
\newblock \emph{Neural computation}, 9\penalty0 (8):\penalty0 1735--1780, 1997.

\bibitem[Choi et~al.(2023)Choi, Jeon, and Lim]{choi2023optimal}
K.~J. Choi, J.~Jeon, and B.~H. Lim.
\newblock Optimal staking and liquid token holding decisions in cryptocurrency
  markets.
\newblock \emph{Available at SSRN 4528742}, 2023.

\bibitem[Xu and Feng(2022)]{xu2022reap}
J.~Xu and Y.~Feng.
\newblock Reap the harvest on blockchain: A survey of yield farming protocols.
\newblock \emph{IEEE Transactions on Network and Service Management},
  20\penalty0 (1):\penalty0 858--869, 2022.

\bibitem[He et~al.(2020)He, Tang, and Wang]{he2020staking}
P.~He, D.~Tang, and J.~Wang.
\newblock Staking pool centralization in proof-of-stake blockchain network.
\newblock \emph{Available at SSRN 3609817}, 2020.

\bibitem[Gersbach et~al.(2022)Gersbach, Mamageishvili, and
  Schneider]{gersbach2022staking}
H.~Gersbach, A.~Mamageishvili, and M.~Schneider.
\newblock Staking pools on blockchains.
\newblock \emph{arXiv preprint arXiv:2203.05838}, 2022.

\bibitem[Tang et~al.(2022)Tang, Song, Zhu, Yuan, Hou, Ji, Tang, and
  Li]{tang2022survey}
Y.~Tang, Z.~Song, Y.~Zhu, H.~Yuan, M.~Hou, J.~Ji, C.~Tang, and J.~Li.
\newblock A survey on machine learning models for financial time series
  forecasting.
\newblock \emph{Neurocomputing}, 512:\penalty0 363--380, 2022.

\bibitem[Soni et~al.(2022)Soni, Tewari, and Krishnan]{soni2022machine}
P.~Soni, Y.~Tewari, and D.~Krishnan.
\newblock Machine learning approaches in stock price prediction: A systematic
  review.
\newblock In \emph{Journal of Physics: Conference Series}, volume 2161, page
  012065. IOP Publishing, 2022.

\bibitem[Kyriazis et~al.(2020)Kyriazis, Papadamou, and
  Corbet]{kyriazis2020systematic}
N.~Kyriazis, S.~Papadamou, and S.~Corbet.
\newblock A systematic review of the bubble dynamics of cryptocurrency prices.
\newblock \emph{Research in International Business and Finance}, 54:\penalty0
  101254, 2020.

\bibitem[Khedr et~al.(2021)Khedr, Arif, El-Bannany, Alhashmi, and
  Sreedharan]{khedr2021cryptocurrency}
A.~M. Khedr, I.~Arif, M.~El-Bannany, S.~M. Alhashmi, and M.~Sreedharan.
\newblock Cryptocurrency price prediction using traditional statistical and
  machine-learning techniques: A survey.
\newblock \emph{Intelligent Systems in Accounting, Finance and Management},
  28\penalty0 (1):\penalty0 3--34, 2021.

\bibitem[Amirzadeh et~al.(2022)Amirzadeh, Nazari, and
  Thiruvady]{amirzadeh2022applying}
R.~Amirzadeh, A.~Nazari, and D.~Thiruvady.
\newblock Applying artificial intelligence in cryptocurrency markets: A survey.
\newblock \emph{Algorithms}, 15\penalty0 (11):\penalty0 428, 2022.

\bibitem[Mudassir et~al.(2020)Mudassir, Bennbaia, Unal, and
  Hammoudeh]{mudassir2020time}
M.~Mudassir, S.~Bennbaia, D.~Unal, and M.~Hammoudeh.
\newblock Time-series forecasting of bitcoin prices using high-dimensional
  features: a machine learning approach.
\newblock \emph{Neural computing and applications}, pages 1--15, 2020.

\bibitem[Hamayel and Owda(2021)]{hamayel2021novel}
M.~J. Hamayel and A.~Y. Owda.
\newblock A novel cryptocurrency price prediction model using gru, lstm and
  bi-lstm machine learning algorithms.
\newblock \emph{AI}, 2\penalty0 (4):\penalty0 477--496, 2021.

\bibitem[Derbentsev et~al.(2021)Derbentsev, Babenko, Khrustalev, Obruch, and
  Khrustalova]{derbentsev2021comparative}
V.~Derbentsev, V.~Babenko, K.~Khrustalev, H.~Obruch, and S.~Khrustalova.
\newblock Comparative performance of machine learning ensemble algorithms for
  forecasting cryptocurrency prices.
\newblock \emph{International Journal of Engineering}, 34\penalty0
  (1):\penalty0 140--148, 2021.

\bibitem[D’Amato et~al.(2022)D’Amato, Levantesi, and Piscopo]{d2022deep}
V.~D’Amato, S.~Levantesi, and G.~Piscopo.
\newblock Deep learning in predicting cryptocurrency volatility.
\newblock \emph{Physica A: Statistical Mechanics and its Applications},
  596:\penalty0 127158, 2022.

\end{thebibliography}

\end{document}